# Handling Missing MRI Input Data in Deep Learning Segmentation of Brain Metastases: A Multi-Center Study

[1,2]Endre Grøvik[*], [3]Darvin Yi[*], [2]Michael Iv, [2]Elizabeth Tong, [1]Line Brennhaug Nilsen, [4]Anna Latysheva, [4]Cathrine Saxhaug, [5]Kari Dolven Jacobsen, [5]Åslaug Helland, [2]Kyrre Eeg Emblem[†], [1]Daniel Rubin[†] and [2]Greg Zaharchuk[†]

**ABSTRACT:** The purpose of this study was to assess the clinical value of a novel DropOut model for detecting and segmenting brain metastases, in which a neural network is trained on four distinct MRI sequences using an input dropout layer, thus simulating the scenario of missing MRI data by training on the full set and all possible subsets of the input data. This retrospective, multi-center study, evaluated 165 patients with brain metastases. A deep learning (DL) based segmentation model for automatic segmentation of brain metastases, named DropOut, was trained on multi-sequence MRI from 100 patients, and validated/tested on 10/55 patients. The segmentation results were compared with the performance of a state-of-the-art DeepLabV3 model. The MR sequences in the training set included pre- and post-gadolinium (Gd) T1-weighted 3D fast spin echo, post-Gd T1-weighted inversion recovery (IR) prepped fast spoiled gradient echo, and 3D fluid attenuated inversion recovery (FLAIR), whereas the test set did not include the IR prepped image-series. The ground truth segmentations were established by experienced neuroradiologists. The results were evaluated using precision, recall, Dice score, and receiver operating characteristics (ROC) curve statistics, while the Wilcoxon rank sum test was used to compare the performance of the two neural networks. The area under the ROC curve (AUC), averaged across all test cases, was 0.989±0.029 for the DropOut model and 0.989±0.023 for the DeepLabV3 model (p=0.62). The DropOut model showed a significantly higher Dice score compared to the DeepLabV3 model (0.795±0.105 vs. 0.774±0.104, p=0.017), and a significantly lower average false positive rate of 3.6/patient vs. 7.0/patient (p<0.001) using a 10mm$^3$ lesion-size limit. The DropOut neural network, trained on all possible combinations of four MRI sequences, may facilitate accurate detection and segmentation of brain metastases on a multi-center basis, even when the test cohort is missing MRI input data.

**Keywords:** Deep Learning, Segmentation, Brain Metastases, Multi-sequence MRI

[*]*Co-first authorship (alphabetic order)*
[†]*These authors contributed equally to this work (alphabetic order)*

[1]*Department for Diagnostic Physics, Oslo University Hospital, Norway;* [2]*Department of Radiology, Stanford University, USA;* [3]*Department of Biomedical Data Science, Stanford University, USA;* [4]*Department of Radiology and Nuclear Medicine, Oslo University Hospital, Norway;* [5]*Department of Oncology, Oslo University Hospital, Norway*

**Corresponding Author:** *Greg Zaharchuk, Department of Radiology, Stanford University, School of Medicine, 1201 Welch Road, Stanford, California 94305-5488, USA. Phone: (650) 736-6172, Fax: (650) 723-9222. Email: gregz@stanford.edu.*

## INTRODUCTION

Advances in artificial intelligence (AI) are suggesting the possibility of new paradigms in healthcare and are particularly well-suited to be adopted by radiologists (1–4). In recent years, there has been significant effort in utilizing the next-generation AI technology, coined *deep learning*, to learn from labeled magnetic resonance imaging (MRI) data (5–7). One key advantage of AI-based radiology is the automatization and standardization of tedious and time-consuming tasks, most clearly exemplified in the tasks surrounding detection and segmentation (8–10). Several deep learning approaches have successfully been developed and tested for automatic segmentation of gliomas (11), thanks in part to the publicly available Brain Tumor Segmentation (BraTS) dataset (12). In recent years, studies have also shown the potential of AI-based segmentation in patient cohorts comprising tumor subtypes, such as brain metastases, which may pose a greater challenge in terms of segmentation performance given their wide range of sizes and multiplicity (13, 14). In a recent study, we trained a fully convolution neural network (CNN) for automatic detection and segmentation of brain metastases using multi-modal MRI (15). While our DL-approach showed high performance, the robustness and clinical utility needs to be challenged in order to fully understand its strengths and limitations. In fact, many AI-based segmentation studies are limited in terms of generalizability in that the algorithms are trained and tested on single-center patient cohorts. In some studies, the training- and test-sets are even limited to a single magnetic field strength, a single vendor, and/or a single scanner for data acquisition. A key step towards understanding the generalizability and clinical value of any deep neural network is by training and testing using real-world multicenter data. Another limitation of these AI-based segmentation networks is that they are trained on a distinct set of MRI contrasts, which limits the use of the networks to sites acquiring the same sequences. However, deep neural networks should be able to handle missing model inputs.

To this end, this work tested a novel AI-based segmentation model, called *DropOut*, in which a neural network with an input dropout layer is trained on the full set of four distinct MRI sequences, as well as every possible subset of the input channels. The DropOut network was tested on a patient cohort missing one of the four input sequences used for training. To evaluate this network's performance, a second neural network was trained and tested using state-of-the-art architecture without applying the DropOut strategy, i.e. only trained on the limited sequences corresponding to those in the test set. We hypothesize that the DropOut model will yield segmentation performance comparable to that of a state-of-the-art segmentation network, while at the same time being robust towards missing input data and allow it to generalize to multicenter MRI data.

## MATERIALS AND METHODS

### Patient Population

This retrospective, multi-center study was approved by our Institutional Review Board. The patient cohort consisted of a total of 165 patients with brain metastases, enrolled from two different hospitals, hereinafter referred to as *'Hospital A'* and *'Hospital B'*. From Hospital A, MRI data from a total of 100 patients were acquired and used for neural network training. A total of 65 patients from Hospital B were used for validation and testing.

Inclusion criteria for the training data included the presence of known or possible metastatic disease (i.e., presence of a primary tumor), no prior surgical or radiation therapy, and the availability of all required MR imaging sequences (see below). Only patients with ≥ 1 metastatic lesion were included. Mild patient motion was not an exclusion criterion. For the validation and test data, we used MRI data from an ongoing clinical study entitled *TREATMENT* (NCT03458455) conducted at Hospital B. To be eligible for inclusion, patients had to receive stereotactic radiosurgery (SRS) for at least one brain metastasis measured at a minimum of 5 mm in one direction, be untreated or progressive after systemic or local therapy, have confirmed non-small-cell lung cancer (NSCLC) or malignant melanoma, be ≥18 years of age; have an Eastern Cooperative Oncology Group performance status score ≤1, and have a life expectancy >6 weeks. Details on the patient cohorts are shown in Table 1.

### Imaging Protocol

For the training set, imaging was performed on both 1.5T (n=7; SIGNA Explorer and TwinSpeed, GE Healthcare, Chicago, IL) and 3T (n=93; Discovery 750 and 750w and SIGNA Architect, GE Healthcare, Chicago, IL, USA; Skyra, Siemens Healthineers, Erlangen, Germany) clinical scanners. The imaging protocol included pre- and post-Gadolinium (Gd) T1-weighted 3D fast spin echo (CUBE/SPACE), post-Gd T1-weighted 3D axial inversion recovery prepped fast spoiled gradient-echo (IR-FSPGR) (BRAVO/MPRAGE), and 3D CUBE/SPACE fluid-attenuated inversion recovery (FLAIR). For Gd-enhanced imaging, a dose of 0.1 mmol/kg body weight of gadobenate dimeglumine (MultiHance, Bracco Diagnostics, Princeton, New Jersey) was intravenously administered.

**Table 1:** Patient demographics

|  | Hospital A | Hospital B |
|---|---|---|
| # of patients | 100 | 65 |
| Gender | 71 F / 29 M | 35 F / 30 M |
| Mean age (range) | 64 (32 – 92) | 65 (32 – 86) |
| **Primary cancer**: | | |
| Lung | 66 | 45 |
| Skin/melanoma | 4 | 20 |
| Breast | 26 | - |
| Genitourinary | 2 | - |
| Gastrointestinal | 2 | - |

**Table 2**: Overview of MRI pulse sequences and key imaging parameters

| Technique | 3D T1 BRAVO | Pre/Post 3D T1 CUBE/SPACE | 3D FLAIR |
|---|---|---|---|
| **Hospital A Data** | | | |
| TR (ms) * | 12.02 / 8.24 | 550 / 602 | 6000 |
| TE (ms)* | 5.05 / 3.24 | 9.54 / 12.72 | 119 / 136 |
| Flip angle* | 20 / 13 | 90 | 90 |
| FOV (mm$^2$) | 240×240 | 250×250 | 240×240 |
| Inversion time (ms) * | 300 / 400 | - | 1880 / 1700 |
| Acquisition matrix | 256×256 | 256×256 | 256×256 |
| Slice thickness (mm) | 1 | 1 | 1 – 1.6 |
| # of slices | 160 | 270 – 320 | 270 – 320 |
| Slice acquisition plane | Axial | Sagittal | Sagittal |
| **Hospital B Data**** | | | |
| TR (ms) | – | 700 | 5000 |
| TE (ms) | – | 12 | 387 |
| Flip angle | – | 120 | 120 |
| FOV (mm$^2$) | – | 230×230 | 230×230 |
| Inversion time (ms) | – | – | 1800 |
| Acquisition matrix | – | 256×256 | 256×256 |
| Slice thickness (mm) | – | 0.9 | 0.9 |
| # of slices | – | 192 | 208 |
| Slice acquisition plane | – | Sagittal | Sagittal |

TR = repetition time, TE = echo time, FOV = field-of-view, BRAVO – T1-weighted inversion recovery prepped fast spoiled gradient-echo, CUBE/SPACE – T1-weigthed fast spin-echo, FLAIR – fluid attenuated inversion recovery.

\* In case of varying parametric values between field strength, '/' notation is given (1.5T / 3T).

** Note that the Hospital B data is missing 3D T1 BRAVO images

For the test set, imaging (n=65) was performed on a clinical 3T Skyra scanner (Siemens Healthineers, Erlangen, Germany). The imaging protocol included pre- and post-Gd T1-weighted 3D fast spin echo (SPACE) and 3D T2-weighted FLAIR. All sequences with key imaging parameters are summarized in Table 2.

**Image pre-processing and segmentation**

For the training data, ground truth segmentations were established by two neuroradiologists with 9 and 3 years of experience by manually delineating and cross-checking regions of interests (ROIs) for every Gd-enhancing metastatic lesion. The lesions were outlined on each slice on the post-Gd 3D T1-weighted IR-FSPGR sequence, with additional guidance from the 3D FLAIR and the post-Gd 3D T1-weighted spin echo data using the OsiriX MD software package (Version 8.0, Geneva, Switzerland).

For the test data, ground truth segmentations of Gd-enhancing metastatic lesions were manually drawn on post-Gd 3D T1-weighted spin echo data by two radiologists with 14 and 5 years of relevant experience. Delineations were performed using the nordicICE software package (NordicNeuroLab, Bergen, Norway).

All imaging series were co-registered into one common anatomical space. For the training data, pre- and post-Gd 3D T1-weighted spin echo data and FLAIR were co-registered to the post-Gd 3D T1-weighted IR-FSPGR, whereas for the test data, the post-Gd 3D T1-weighted spin echo images was used as the reference series for the pre-Gd 3D T1-weighted spin echo data and FLAIR. Prior to network training, a defacing procedure was applied to anonymize all imaging data using an in-house algorithm (MATLAB R2017a version 9.2.0, MathWorks Inc. Natick, MA).

**Neural Network Details**

The neural networks used in this study were based on the DeepLab V3 architecture (16), and the modifications and training strategies are detailed in a recent work (17). In particular, this study utilized and trained a 'input-level integration dropout' network, referred to as the DropOut model (Figure 1). This model was trained on patients from Hospital A, using 5 slices from the four aforementioned pulse sequences as input. These were all stacked together in the color channel, resulting in an image tensor of shape 256x256x20 as the model input. The network was trained by utilizing a pulse-sequence level dropout; replacing the full 5 slices of any given pulse sequence with an empty tensor of 0's during training, thus enabling the network to handle missing MRI pulse sequence input during inference. This yields a network trained on a single data center with a superset of pulse sequences to what may be used in practice.

A key mathematical note is that when pulse sequences are censored, the remaining sequences must be multiplied by a constant value to keep the pixel statistics relatively constant. For example, if a single pulse sequence is censored, the remaining three pulse sequence inputs are multiplied by a value of 4/3. If two pulse sequences are censored, the remaining pulse sequences are multiplied by

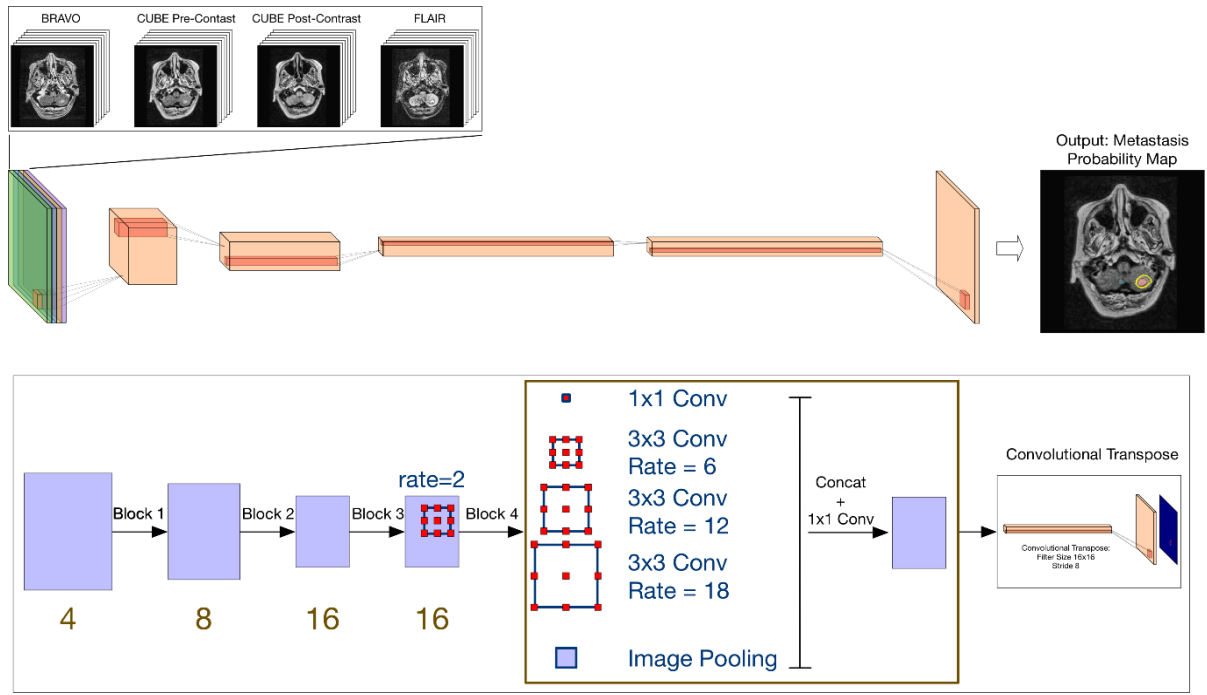

*Figure 1:* Diagram showing the DeepLab V3 based segmentation network used in this study. Five contiguous axial slices of each of the four pulse sequences (BRAVO, pre- and post-Gd CUBE, and T2-weighted FLAIR) are concatenated in the color-channel dimension to create an input tensor with channel dimension 20. This is fed into a DeepLab V3 based network to predict the segmentation on the center slice.

2 (= 4/2). Thus, when inference was performed on patients from Hospital B, having only three of the four total pulse sequences, the corresponding slices of the 'missing' pulse sequence was defined as an empty 0-tensor, while the available three pulse sequences were multiplied by 4/3.

To evaluate the segmentation performance of the DropOut model, a second neural network was trained and tested using the state-of-the-art DeepLab V3 architecture without applying the input-level dropout strategy, and only trained on the complete set of sequences matching that of the test set from Hospital B (i.e. excluding the post-Gd 3D T1-weighted IR-FSPGR sequence).

All patients from Hospital A were used for training, while the patients from Hospital B were divided into training and validation sets, giving a final breakdown 100 training cases, 10 validation cases, and 55 test cases. All training was done on a system with two NVIDIA GTX 1080Ti GPUs.

**Statistical Analysis**

The networks' ability to detect metastatic lesions on a voxel-by-voxel basis was evaluated using receiver operating characteristic (ROC) curve statistics, measuring the area under the ROC curve (AUC) for each patient in the test set. Further, the optimal probability threshold for including a voxel within the metastatic lesion was determined using the Youden index from the ROC statistics on the validation set. Using this threshold, the networks segmentation performance was further evaluated by estimating the *precision-* and *recall*-values, as well as the Dice similarity score (also known as the F1 score).

In addition to evaluating the neural networks' ability to detect metastatic tissue voxels, performance was also evaluated on a lesion-by-lesion basis by calculating the number of false positive (FP) per case. The FP was determined by multiplying the ground truth maps and the thresholded probability maps and counting the number of overlapping objects in the resulting binary image using a connecting component approach. Voxels were considered connected if their edges or corner touch. The number of FP was determined both without any size criterion, as well as only considering objects ≥ 10 $mm^3$ (roughly 2 mm in linear dimension) as a detected lesion.

Finally, the Wilcoxon rank sum test was used to compare the performance of the DropOut model and the DeepLabV3 network. A statistical significance level of 5% was used. All statistical analyses were performed using MATLAB R2017a version 9.2.0 (MathWorks Inc. Natick, MA).

**RESULTS**

The total time used for training was approximately 20 hours for both the DropOut model and the DeepLabV3 network. For processing a test case using the DropOut model, the forward pass on a system with two NVIDIA GTX 1080Ti GPUs took approximately 250 ms per slice.

Figure 2 show six example cases demonstrating the resulting probability maps, as well as maps representing the performance in terms of true positive, false positive, and false negative, as an overlay on the post-Gd 3D T1-weighted spin echo image series.

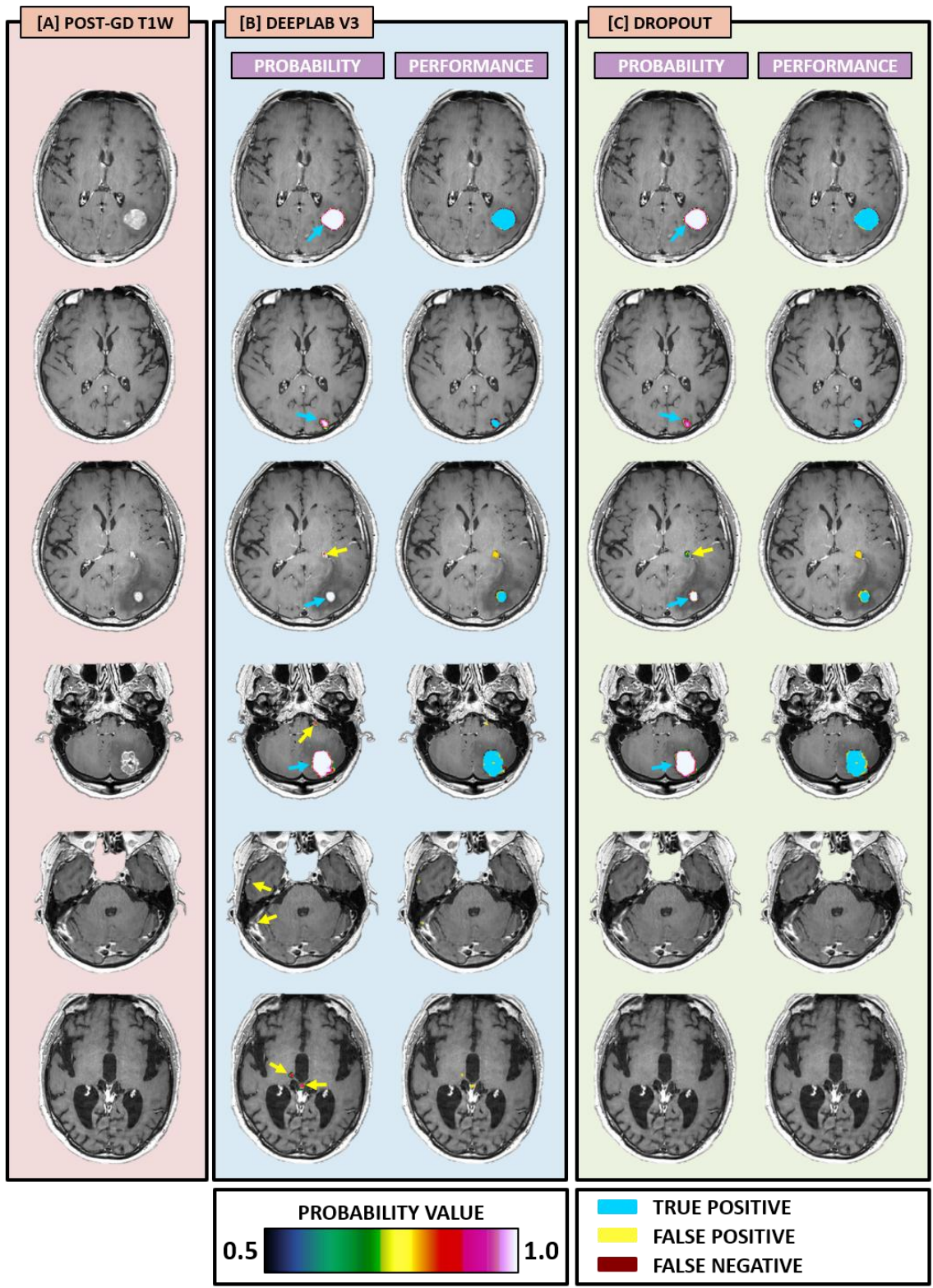

*Figure 2:* Examples in representative test set cases showing the segmentation predictions from the [B] DeepLab V3 model and the [C] DropOut method. The image mosaic shows the predictions as probability maps (voxel-wise ranging from 0 to 1 as indicated by the color bar) and performance maps classified as true negative, false positive, and false negative as specified by the color code. All maps are shown as overlays on a post-Gd 3D T1-weighted image series [A]. The cases shown here are [first row] a 65-year-old female with malignant melanoma, [second row], 73-year-old male with non-small cell lung cancer (NSCLC), [third row] 66-year-old male with NSCLC, [fourth row] 44-year-old female with NSCLC, [fifth row] 64-year-old female with NSCLC, and [sixth row] 73-year-old male with NSCLC. The blue arrows indicate true positive lesions, while yellow arrows indicate false positive lesions. Note that in the bottom three cases, the DeepLab V3 returns several false positive lesions which are not reported by the DropOut method, thus reflecting the results indicating a superior performance on false positive rate by the DropOut method.

DropOut model and the DeepLabV3 network are summarized in Table 3. Both the DropOut model and the DeepLab v3 network show a high voxel-wise detection accuracy, yielding an AUC, averaged across all test cases, of 0.989 ± 0.029 and 0.989 ± 0.029 (NS, p=0.620), respectively (Figure 3).

Based on the ROC analysis on the validation set, an optimal probability threshold for including a voxel as a metastasis was set to 0.76 for the DropOut model, and 0.87 for the DeepLab V3. Using these thresholds, the DropOut model demonstrated a significantly higher Dice score (0.795 ± 0.104) compared to the DeepLab network (0.774 ± 0.104) (p=0.017). The average recall and precision values were also higher for the DropOut model, but this difference was not statistically significant (Table 2).

On a per-lesion basis, and without any lesion-size limit, the DropOut model showed an average FP of 12.3/patient, which was significantly lower that the DeepLabV3 network (26.3/patient, p<0.001). By applying a lesion-size limit of 10 $mm^3$, the DropOut model demonstrated an average FP of 3.6/patient, also significantly lower that the DeepLabV3 (7.0/patient, p<0.001) (Figure 4).

**DISCUSSION**

Detection and segmentation of brain metastases on radiographic images sets the basis for clinical decision making and patient management. Precise segmentation is crucial for several steps in the clinical workflow such as treatment decision, radiation planning and assessing treatment response, and must therefore be performed with the utmost accuracy. Considering that the value of volumetric measurements of enhancing brain lesions are increasingly discussed (18), future manual detection and segmentation pose a tedious and time-consuming challenge, particularly with the growing use of multi-sequence 3D imaging.

In this study, we demonstrated the clinical value of a novel DropOut model for automatic detection and segmentation of brain metastases. This neural network has a unique advantage over other segmentation networks because it uses an input dropout layer; trained on a full set of four MRI input sequences, as well as every possible reduced subset of the input channels, thus simulating the scenario of missing MRI data during training.

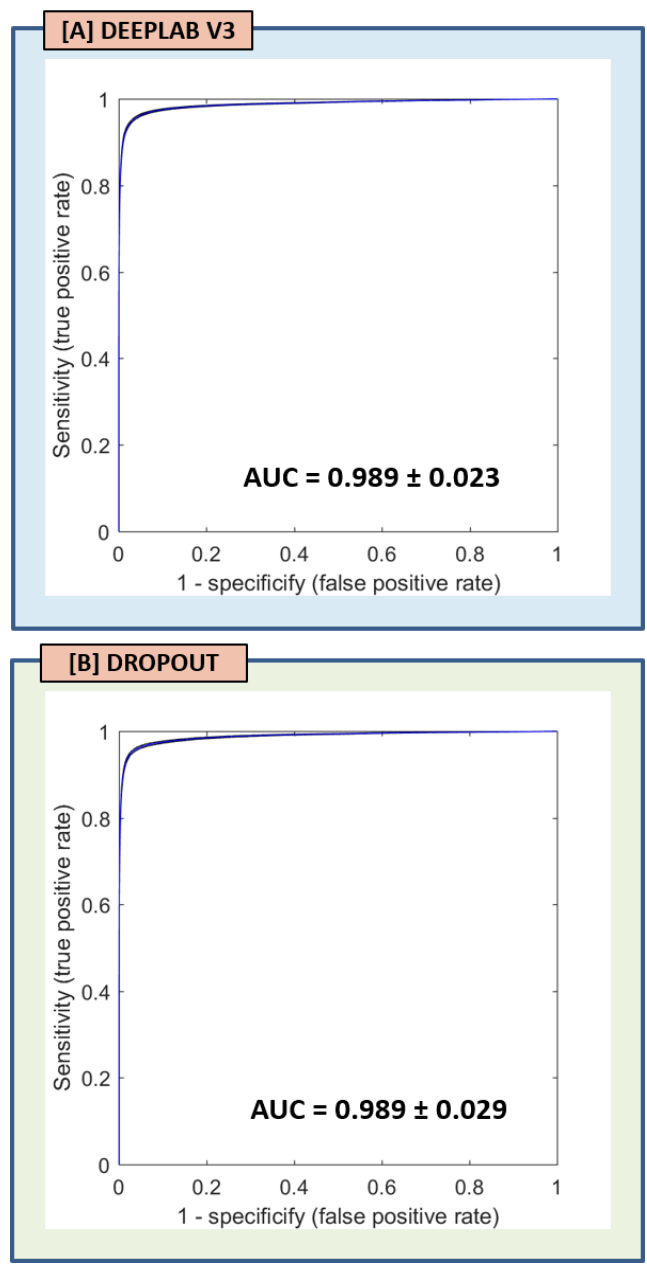

**Figure 3:** ROC curves with 95% confidence interval (shaded area) averaged across all 55 test cases for the [A] DeepLab V3 model and the [B] DropOut method. The area under the ROC curve was 0.989 (ranging from 0.896 to 1.000) for the DeepLab V3 model, and 0.989 (ranging from 0.845 to 1.000) for the DropOut model.

**Table 3: Detection accuracy and segmentation performance**

| Network | AUC ROC | DICE | Recall | Precision | FP (no size limit) | FP (10 $mm^3$ size limit) |
|---|---|---|---|---|---|---|
| DeepLab V3 | 0.989±0.023 | 0.774±0.104 | 0.631±0.208 | 0.722±0.206 | 26.3±17.2 | 7.0±5.3 |
| DropOut | 0.989±0.029 | 0.795±0.105 | 0.671±0.262 | 0.790±0.158 | 12.3±10.2 | 3.6±4.1 |
| P-value | 0.620 | **0.017** | 0.167 | 0.095 | **<0.001** | **<0.001** |

All metrics except AUC ROC were estimated using a probability threshold of 0.87 for the DeepLab V3 model, and 0.76 for the DropOut model

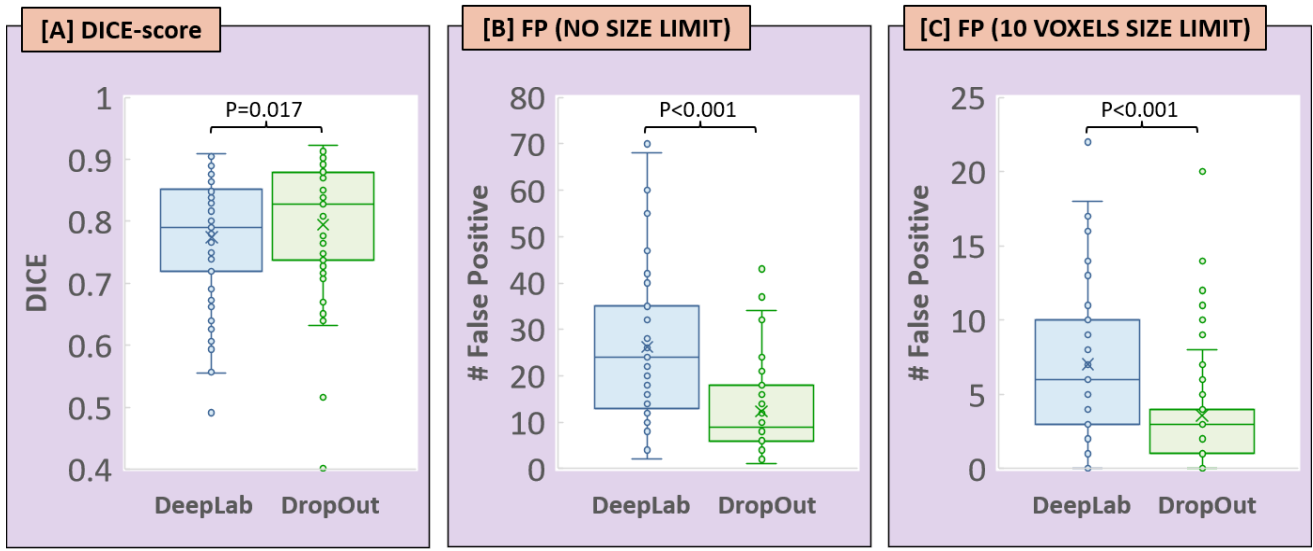

*Figure 4: Boxplot showing the resulting [A] Dice score, as well as the number of false positive [B] without and [C] with a lesion-size limit of 10 mm$^3$, as determined from the segmentation probability maps produced by the DeepLab model (blue) and the in-house DropOut method (green).*

Consequently, the resulting DropOut model can return a model output (probability map) regardless of missing sequences. The accuracy of the DropOut model in detecting metastatic voxels in the brain, as measured by the AUC, is equivalent to that of the state-of-the-art DeepLab V3 neural network trained on the specific subset of sequences in our test set. However, our results indicate that the proposed model is superior to the DeepLab V3 in terms of segmentation performance, as measured by the Dice score, and at the same time returning significantly fewer FP. This generalization was also achieved despite differences in the patient demographics between the training and test sets, with more frequent representation of lung and melanoma metastases in the test set. Finally, we would like to emphasize that the DropOut method does not require re-training for another subset of imaging sequences that might be acquired in another institution.

The neural networks used in this study were based on the DeepLab V3 architecture, which is currently considered as one of the most robust neural networks for image-based semantic segmentation. The key difference of the DeepLab V3 compared with other relevant architectures is its reliance on atrous (or dilated) convolutions. By using atrous convolutional layers, our network has a very large receptive field, thereby incorporating greater spatial context. This approach may be key to enabling the network to identify local features as well as global contexts, i.e. identifying brain regions, which could enhance the network's decision-making process on similar local features.

In our study, the networks' performance was tested on multi-center data, representing an essential step towards understanding the generalizability and clinical value of the proposed neural network. In this sense, it represents a logical extension of our prior single-center study on this topic ((15). No previous studies have evaluated deep learning for brain metastasis detection using multi-center data. Other single-center studies, such as Liu et al. (13) and Charron et al. (14), have recently shown that deep neural networks can detect and segment brain metastases with high accuracy and performance, reporting results comparable to that of the current study. The latter study also demonstrated that a deep neural network trained on multi-sequence MRI data outperformed single contrast networks.

In general, the two most commonly used MRI sequences for assessing brain metastases are post-Gd T1-weighted and T2-weighted FLAIR. The post-Gd 3D T1-weighted, high-resolution isotropic sequence is most crucial (19) and can be acquired by fast spin-echo or gradient-echo techniques. The 3D T1-weighted gradient-echo sequences (e.g. IR-FSPGR, BRAVO, MPRAGE) are broadly used because they create isotropic T1-weighted images with excellent grey-white matter differentiation but are limited by lower contrast conspicuity and a lack of blood vessel suppression. The 3D fast spin-echo techniques (e.g. CUBE, SPACE, VISTA) are relatively newer techniques optimized for isotropic high-resolution 3D imaging of T1-weighted, T2-weighted, or FLAIR images; and have the advantage of blood vessel suppression. For this study, post-Gd T1-weighted 3D fast spin-echo, pre- and post-Gd T1-weighted 3D axial IR-FSPGR, and 3D FLAIR sequences were used as input to train the neural network. While these sequences are widely used for imaging brain metastases, they are not compulsory. Variations in sequences and acquisition parameters among different institutions also frequently are present. For instance, 2D FLAIR (with thicker slice and non-isotropic voxels) may be acquired instead of 3D FLAIR. In clinical practice, it is also not unusual to omit sequences owing to patients' safety or comfort. Therefore, it is imperative to design a robust and versatile neural network that can accommodate missing sequences while maintaining good performance.

While this study shows a high accuracy using the DropOut model for detecting and segmenting brain metastases, the results should be interpreted in light of the limited sample size and the homogeneity of the test cohort. Patients included in the test set were all scheduled for SRS, which generally presents with fewer and larger metastases, which in turn may be easier for the network to predict. This hypothesis is supported by observations made in our previous study, in which the tested neural network showed higher accuracy in patients with 3 or less metastases compared to patients with >3 metastases (15). However, a total of nine patients in the current test set presented with >3 small metastases, for which the DropOut model still demonstrated a high accuracy and performance, equivalent to the average metrics for all lesions.

**In conclusion, this study demonstrates that the DropOut method, utilizing a pulse sequence level dropout layer, thus being trained on all possible combinations of multiple MRI sequences, can detect and segment brain metastases with high accuracy, even when the test cohort is missing input data. This is likely of value for generalizing deep learning models for use in multiple different imaging sites.**